\documentclass{article}
\usepackage{amsmath,amssymb,pxfonts}
\title{Conservative delta hedging under transaction costs}
\author{Masaaki Fukasawa \\ 
{\small Department of Mathematics, Osaka University}}
\date{}
\newtheorem{defn}{Definition}
\newtheorem{prop}{Proposition}
\newtheorem{thm}{Theorem}
\newtheorem{assu}{Condition}
\newtheorem{lem}{Lemma}
\begin{document}
\maketitle
\begin{abstract}
Explicit robust hedging strategies 
for convex or concave payoffs
under a continuous semimartingale model with
uncertainty and small transaction costs are constructed.
In an asymptotic sense,
the upper and lower bounds of the cumulative volatility 
enable us to super-hedge convex and concave payoffs
respectively.
The idea is a combination of 
Mykland's conservative delta hedging
 and Leland's enlarging volatility.
We use a specific sequence of stopping times as rebalancing dates,
which can be superior to equidistant one even when there is no
model uncertainty. 
A central limit theorem for the super-hedging error as the coefficient of 
linear transaction costs tends to zero is proved.
The mean squared error is also studied.
\end{abstract}

\begin{flushleft}


\end{flushleft}

\section{Introduction}

Hedging an European option by dynamically trading the underlying asset
is the very basic of mathematical finance.
It is still a practical problem since 
liquid option markets do not exist for all kinds of payoffs written on
all kinds of assets.
In particular, 
a dynamic hedging strategy is required for option contracts with long
time-to-maturity.
To deal with a long period of hedging, problematic is the model uncertainty of
the underlying asset price process.
Super-hedging under
the model uncertainty has been recently paid much attention in the literature
of mathematical finance; 
Avellaneda, Levy and Paras~\cite{Ave}, Lyons~\cite{Lyons},
Mykland~\cite{Mykland2000,Mykland2003b,Mykland2003},
Denis and Martini~\cite{DM}, Cont~\cite{Cont}, 
Peng~\cite{Penga,Pengb}, Soner, Touzi and Zhang~\cite{STZ} among others.
Another issue of dynamic hedging is that 
rebalancing portfolio should be done discretely and 
associated with small but non-zero transaction costs;
see e.g., Kabanov and Safarian~\cite{Kabanov},
Bouchard and Touzi~\cite{BouchardTouzi},
Leland~\cite{Leland}, Gamys and Kabanov~\cite{GamysKabanov},
Denis and Kabanov~\cite{DenisKabanov}, and
Fukasawa~\cite{aap,safa}.

The aim of this paper is to present a practical strategy which takes
both the uncertainty and the transaction costs into account.
We employ the familiar Black-Scholes pricing function and 
the delta hedging strategy
with specific choices of volatility parameter and rebalancing dates.
We suppose that the underlying asset price process and the
zero-coupon bond price process are positive continuous semimartingale,
that the option payoff is either convex or concave, 
and that the transaction costs follow a linear model. 

To cope with the model uncertainty, we adopt
a continuous-time trading strategy which is a variant of 
Mykland~\cite{Mykland2000}'s conservative delta hedging.
The same idea was also introduced by
Carr and Lee~\cite{CL}, where emphasis was put on its application to 
model-free hedging of variance options.
We rely only on bounds on the cumulative volatility, that is,
the quadratic variation  of the log price process up to the maturity of
the option.
It is reasonable to suppose the availability of such bounds 
because the volatility is typically observed to be
mean-reverting, so that the cumulative volatility across a long period
is naturally supposed to be stable and predictable. 
This is a different approach from one given by 
Avellaneda, Levy and Paras~\cite{Ave}
and its extensions, which relies on bounds on not the cumulative
volatility but the volatility itself.
Our approach turns out to be more efficient
as long as considering an European option written on one asset.
Note that a bound on the volatility implies in particular 
a bound on the cumulative
volatility.
Since we are dealing with a general continuous semimartingale,
it is clear that without any bounds on the cumulative volatility, 
we cannot have anything 
better than the trivial buy-and-hold super-hedging strategy. 

To take the transaction costs into consideration, 
we construct a discretized version of the conservative 
delta hedging strategy inspired by
Leland~\cite{Leland}'s idea of
enlarging volatility for the Black-Scholes model.
The main result of this paper is the stable convergence of the discrete
hedging strategy as the coefficient of the transaction costs converges
to $0$. 
The mean squared error is also investigated.
In particular, the result implies in an asymptotic sense 
that by following this strategy, 
we can super-hedge convex and concave payoffs under
the upper and lower bounds of the cumulative volatility respectively.
Note that without employing such an asymptotic framework, 
we do not have anything 
better than the trivial buy-and-hold strategy,
as shown in Bouchard and Touzi~\cite{BouchardTouzi}.
An alternative approach is 
the framework of pricing and partial hedging under loss constraints.
See Bouchard and Dang~\cite{BouchardDang}.
The advantage of our asymptotic approach  is the availability of
the explicit expressions of the strategy and the corresponding 
loss distribution.
Not only that everything is explicit but also 
that everything is  written in terms of the Black-Scholes greeks,
 so it is quite easy to implement our strategy in financial practice.
Our specification of rebalancing dates
plays a crucial role; it is not from technical convenience 
but reflecting the structure of the transaction costs.
Our rebalancing dates 
are not deterministic, so that the convergence results
are new even under the Black-Scholes model
that is a special case of our framework.

In Section~2, we describe the continuous-time strategy under no
transaction costs. In Section~3, we construct the discrete hedging
strategy and present the main result.
In Section~4, we treat the mean squared error of the strategy.
In Section~5, we compare the mean square errors of our strategy and
Leland's strategy under the Black-Scholes model with no model uncertainty.

\section{Conservative delta hedging}
Here we study the super-hedging problem
under no transaction costs.
Let $(\Omega, \mathcal{F}, \mathbb{P}, \{\mathcal{F}_t\})$ be 
a filtered probability space satisfying the usual assumptions.
Let $S^1$ be a one-dimensional positive continuous 
$\{\mathcal{F}_t\}$-semimartingale standing for 
a tradable asset price process.  
We consider hedging an European claim $f(S^1_T)$ by dynamically trading the
underlying asset and the risk-free zero coupon bond with the same
maturity $T$, where  $T > 0$ is fixed and 
$f$ is a convex or concave
function whose right and left derivatives $f^\prime_{\pm}$
are 
of polynomial growth.
Here we say a function $g$ is of polynomial growth if there exists 
$p >0$ such that
\begin{equation*}
\sup_{s \in (0,\infty)} \frac{|g(s)|}{s^p + s^{-p}} < \infty.
\end{equation*}
We denote by $S^0$ the price process of 
the zero coupon bond with maturity $T$.
By definition we have $S^0_T = 1$. 
We extend $S^0_t = 1$ for $t > T$ and suppose that
$S^0$ also is a positive continuous $\{\mathcal{F}_t\}$-semimartingale.
Denote by $\tilde{S}=S^1/S^0$ the discounted price process.
Naturally we suppose that the path of $(S^1,S^0)$ is observable,
 so that in particular
we know the quadratic variation 
$\langle \log(\tilde{S}) \rangle_t$ at time $t$ for any $t \in [0,T]$.

Now we describe a hedging strategy under no transaction costs
which is a variant of one given in Mykland~\cite{Mykland2000}
and essentially the same as one given in Carr and Lee~\cite{CL}.
 Define a function 
$P:\mathbb{R}_+ \times \mathbb{R} \times \mathbb{R}_+ \to \mathbb{R}$
as 
\begin{equation*}
P(S,R,\Sigma) = \exp\{-R\}\int_{-\infty}^{\infty}
f(S\exp\{R-\Sigma/2 + \sqrt{\Sigma}z\})\phi(z)\mathrm{d}z,
\end{equation*}
where $\phi$ is the standard normal density.
Notice that $P(S^1_0,rT,\sigma^2T)$ is the price the European option with
payoff function $f$ under the Black-Scholes model with volatility
$|\sigma|$ and risk-free rate $r$ for any constants
$\sigma, r \in \mathbb{R}$.
What plays an essential role is the following partial differential
equations for $P$, which can be directly checked:
\begin{equation*}
\frac{\partial P}{\partial \Sigma}
= \frac{1}{2}S^2 \frac{\partial^2 P}{\partial S^2}, \ \ 
\frac{\partial P}{\partial R} = 
S \frac{\partial P}{\partial S} - P, \ \ 
\frac{\partial^2 P}{\partial R \partial S} = S \frac{\partial^2
P}{\partial S^2}, \ \ 
\frac{\partial^2 P}{\partial R^2} = S^2\frac{\partial^2P}{\partial S^2}-
\frac{\partial P}{\partial R}
\end{equation*}
on $(0,\infty) \times \mathbb{R} \times (0, \infty)$
 with boundary condition $P(S,0,0)=f(S)$.
Notice that if $f$ is convex or concave, then so is $P$ in $S$, 
which implies the monotonicity of $P$ in $\Sigma$.
For a given constant $\hat{\Sigma} > 0$, set
\begin{equation*}
R_t = - \log(S^0_t), \ \ 
\Sigma_t = \hat{\Sigma}- \langle \log(\tilde{S}) \rangle_t, \ \ 
\tau = \inf\{t > 0: \langle  \log(\tilde{S}) \rangle_t \geq
 \hat{\Sigma}\}
\end{equation*}
and consider the portfolio strategy defined as
\begin{equation*}
\Pi_t = \frac{\partial P}{\partial S}(S^1_t,R_t,
\Sigma_t), \ \ 
\Pi^0_t = (P(S^1_t,R_t,\Sigma_t) 
- \Pi_t S^1_t)/S^0_t
\end{equation*}
for $t \in [0,\tau)$ and 
$(\Pi_t, \Pi^0_t) = (a,b)$ for $t \geq \tau$, where $a$ and $b$
can be chosen  any
 $\mathcal{F}_{\tau}$-measurable random variables
      satisfying 
$a \in  [f^\prime_-(\tilde{S}_{\tau}),
 f^\prime_+(\tilde{S}_{\tau})]$ 
and $b=f(\tilde{S}_{\tau})-a\tilde{S}_{\tau}$.
The following theorem is an immediate result of the above
partial differential equations
 with the aid of It$\hat{\text{o}}$'s formula.

\begin{prop}
The portfolio strategy
 $(\Pi,\Pi^0)$ is self-financing up to
$\tau$,
that is, 
\begin{equation*}
\begin{split}
P(S^1_t,R_t,\Sigma_t) &= \Pi_t S^1_t + \Pi^0_t S^0_t \\
& = 
P(S^1_0,R_0,\Sigma_0) + \int_0^t \Pi_u \mathrm{d}S^1_u
 + \int_0^t \Pi^0_u \mathrm{d}S^0_u
\end{split}
\end{equation*}
for any $t \in [0,\tau]$.
Moreover,
\begin{itemize}
\item  
if $f$ is convex, then
 $(\Pi,\Pi^0)$ is a super-hedging strategy 
for the payoff $f(S^1_T)$  on the set 
$\{\hat{\Sigma} \geq \langle \log(\tilde{S}) \rangle_T\}$, that is,
we have that
\begin{equation*}
P(S^1_T,R_T,\Sigma_T) = P(S^1_T,0,\Sigma_T) \geq f(S^1_T), \ \ 
\tau \geq T
\end{equation*}
  on set 
$\{\hat{\Sigma} \geq \langle \log(\tilde{S}) \rangle_T\}$.
\item 
if $f$ is concave, then
$(\Pi,\Pi^0)$ is a super-hedging strategy 
for the payoff $f(S^1_T)$  on the set 
$\{\hat{\Sigma} \leq \langle \log(\tilde{S}) \rangle_T\}$, that is,
we have that 
\begin{equation*}
P(S^1_{\tau},R_{\tau},\Sigma_{\tau}) = 
S^0_{\tau}f(\tilde{S}_{\tau}) = a S^1_{\tau} + bS^0_{\tau},
\ \ T \geq \tau,
 \ \ 
aS^1_T + bS^0_T \geq f(S^1_T)
\end{equation*}
on the set $\{\hat{\Sigma} \leq \langle \log(\tilde{S}) \rangle_T\}$.
\end{itemize}
\end{prop}
This strategy requires only a suitable specification of $\hat{\Sigma}$
at time $0$.
We do not specify the detail of the dynamics of $(S^1,S^0)$ 
other than its continuity.
The strategy is efficient in the sense that it becomes the perfect
hedging strategy  in the case 
 $\langle \log (\tilde{S}) \rangle_T = \hat{\Sigma}$
a.s..
The uncertainty of the whole dynamics reduces to that 
 of the cumulative volatility of $\tilde{S}$.
The cumulative volatility, that is, the quadratic variation of 
$\log(\tilde{S})$ is
typically observed to be persistent and mean-reverting, so that
it is reasonable to suppose the availability of a reliable 
prediction interval of 
$\langle \log (\tilde{S}) \rangle_T$ at time $0$ based on the past
price behavior.
Especially for the case that $T$ is large, 
which we have in mind as a motivating example,
we may presume an averaging
effect of long-run volatility.

Outside the specified region 
$\{\langle \log(\tilde{S})\rangle_T \geq \hat{\Sigma}\}$
(resp.~$\{\langle \log(\tilde{S})\rangle_T \leq \hat{\Sigma}\}$),
the proposed strategy fails to super-hedge.
The loss at the maturity is given as $ f(S^1_T) - f(\tilde{S}_\tau) $ 
(resp.~$f(S^1_T)-P(S^1_T,0,\Sigma_T)$)  for convex (resp. concave) payoff.
Letting $\hat{\Sigma} \to \infty$ (resp.~$\hat{\Sigma} \to 0$), 
$P(S_0,R_0,\Sigma_0)$ converges to the cost for the trivial buy-and-hold
hedging strategy.
The larger (resp.~smaller) $\hat{\Sigma}$, the more conservative 
strategy at the expense of requiring  the larger initial cost.
The optimal choice of $\hat{\Sigma}$ depends on hedger's forecast of
the distribution of $\langle \log(\tilde{S})\rangle_T$ and risk preference.
See Mykland~\cite{Mykland2003b} for further discussion.

The difference between our strategy
 and the one given in Mykland~\cite{Mykland2000,Mykland2003b,Mykland2003}
is that ours involves dynamic trading of the zero coupon bond and
we do not suppose any bounds on the cumulative interest rate.
Nevertheless, our super-hedging price $P(S^1_0,R_0,\Sigma_0)$
is less than or equal to that
given in Mykland~\cite{Mykland2000,Mykland2003} 
under the same volatility bounds at least if $S^0$ is of bounded
variation.
The point of this strategy is to utilize 
$\langle \log(\tilde{S}) \rangle_t$ when defining
the delta $\Pi_t$.
A bound on the cumulative volatility does not work by itself as shown
by El Karoui, Jeanblanc and Shreve~\cite{EJS}.

The approach introduced by Avellaneda, Levy and Paras~\cite{Ave}
is based on bounds on not the cumulative volatility but the
volatility itself and does not utilize the available information 
$\langle \log(\tilde{S}) \rangle_t$ at all.
In particular it cannot exploit the averaging effect of
long-run volatility.
As a result, starting the same initial value of portfolio,
our strategy covers a wider class of models at least when considering
convex or concave European payoffs.

The price process $(S^1,S^0)$ are observed discretely in practice, 
so that the estimation error of $\langle \log(\tilde{S})\rangle_t$
might be taken into account. Nevertheless, the observation of the prices
is naturally supposed to be
 much more frequent than rebalancing portfolio, so that
we may neglect this estimation error in this study.
See e.g. Fukasawa~\cite{spa} for the estimation error of
$\langle \log(\tilde{S}) \rangle_t$.

\section{Discrete hedging under transaction costs}
The use of the Black-Scholes delta
 would be convenient in
financial practice.
We believe that the above elementary result is important 
from the viewpoint of risk
management. 
To make it more practical, now we consider its discretized version
under transaction costs.
The main source of the transaction costs is the bid-ask spread that is
usually approximated by a linear model.
Namely  we suppose that the loss in
rebalancing portfolio from
$\Pi_{t-} S^1_t + \Pi^0_{t-}S^0_t$ to 
$\Pi_{t} S^1_t + \Pi^0_{t}S^0_t$ is proportional to $|\Delta \Pi_t|S^1_t$.
Note that the self-financing condition requires
$|\Delta \Pi_t|S^1_t = |\Delta_t\Pi^0_t|S^0_t$ under no transaction costs.
Further, we suppose that the coefficient of these linear transaction costs is
``small''. This motivates us to study the asymptotic behavior
of the discrete hedging under the
 transaction costs $\kappa|\Delta \Pi_t| S^1_t$ as $\kappa \to 0$.
We can expect that the limit distribution of the hedging errors 
as $\kappa \to 0$ serves as a
reasonable approximation of the error distribution with 
a fixed ``small''  $\kappa$.

Here we construct our discrete hedging strategy.
The idea is a modification of Leland's strategy of enlarging volatility
for the Black-Scholes model.
Given a constant $\hat{\Sigma} >0$, set
\begin{equation*}
\Sigma^{\pm \alpha}_t = (1 \pm 2/\alpha)\Sigma_t = 
(1 \pm 2/\alpha)(\hat{\Sigma} - \langle \log(\tilde{S}) \rangle_t)
\end{equation*}
and define a portfolio strategy as
\begin{equation*}
\Pi^{\pm\alpha}_t = 
\frac{\partial P}{\partial S}(S^1_t,R_t,\Sigma^{\pm \alpha}_t), \ \ 
\Pi^{0,\pm \alpha}_t = 
(P(S^1_t,R_t,\Sigma^{\pm \alpha}_t) - \Pi^{\pm \alpha}_t S^1_t)/S^0_t
\end{equation*}
for $t \in [0,\tau)$ and 
$(\Pi^{\pm \alpha}_t, \Pi^{0,\pm \alpha}_t) = (a,b)$ for $t \geq \tau$,
where $P$, $R_t$, $\Sigma_t$, $a$, $b$ are the same as before.
Here $\alpha$ is an arbitrary positive constant at this point
and we use $+\alpha$ if the payoff $f$ is convex and use 
$-\alpha$ if it is concave.
In the sequel, $\pm \alpha$ should always be understood as
$+ \alpha$ or $-\alpha$ if $f$ is convex or concave respectively.
For the latter case we assume $\alpha > 2$.
Notice that as $\alpha \to \infty$, 
$\Sigma^{\pm \alpha}_t \to \Sigma_t$, so that $\alpha$ controls 
how much the volatility is enlarged or shrunk.
We construct a strategy which is asymptotically super-hedging for any 
$\alpha$ and our main result relates $\alpha$ to the asymptotic
distribution of the associated hedging error.
As discussed in Section~5,  $\alpha$ will be determined according to 
hedger's attitude to risk in light of our main result.

This way of enlarging or shrinking volatility is different from
Leland's way. 
With suitable choice of rebalancing dates, 
this modification enables us to work
 beyond the Black-Scholes model. Moreover, 
it turns out to be more
 efficient than Leland's strategy even under the Black-Scholes model
in some sense as we see in Section~5.
While Leland used the equi-distant partition of $[0,T]$ as
rebalancing dates, we use the following 
 sequence of stopping times:
\begin{equation} \label{dates}
\tau^\kappa_0 = 0, \ \ 
\tau^\kappa_{j+1} = \inf\{ t > \tau^\kappa_j;
|\Pi^{\pm \alpha}_t - \Pi^{\pm \alpha}_{\tau^\kappa_j}| \geq \alpha \kappa
\tilde{S}_{\tau^\kappa_j}|\Gamma^{\pm \alpha }_{\tau^\kappa_j}|
\},
\end{equation}
where
\begin{equation*}
\Gamma^{\pm \alpha}_u = 
S^0_u \frac{\partial^2 P}{\partial S^2}(S^1_u,R_u,\Sigma^{\pm \alpha}_u)
= \frac{\partial^2 P}{\partial S^2}(\tilde{S}_u,0,\Sigma^{\pm \alpha}_u).
\end{equation*}
To ensure $\tau^\kappa_j < \tau^\kappa_{j+1}$ a.s. for each $j$, 
we put the following condition:
\begin{assu} \label{scon}
For all 
$(S,\Sigma) \in (0,\infty)\times (0,\infty)$,
\begin{equation*}
\left|
\frac{\partial^2 P}{\partial S^2}(S,0,\Sigma) 
\right| > 0.
\end{equation*}
\end{assu}
Note that we are assuming that $f$ is convex or concave, so we have already
\begin{equation*}
\frac{\partial^2 P}{\partial S^2}(S,R,\Sigma) 
\geq  0 \text{ or }
\frac{\partial^2 P}{\partial S^2}(S,R,\Sigma) 
\leq  0.
\end{equation*}
Therefore, 
Condition~\ref{scon} is a fairly mild condition from practical point of
view. This is violated
if $f^\prime_+$ is continuous singular.

We set the price of the option with payoff $f(S^1_T)$
to be 
\begin{equation} \label{price}
P(S^1_0,R_0,\Sigma^{\pm\alpha}_0) +  \kappa|\Pi^{\pm
 \alpha}_0|\tilde{S}_0
\end{equation}
and follow the delta strategy $\Pi^{\pm \alpha}$
 using $\{\tau^\kappa_j\}$ as
rebalancing dates in the self-financing manner up to $\tau$.
More precisely, we define $\hat{\Pi}^{0,\pm\alpha,\kappa}$ recursively as
\begin{equation*}
\begin{split}
&\hat{\Pi}^{0,\pm\alpha,\kappa}_0 = 
\Pi^{0,\pm \alpha}_0,\\ 
&(\Pi^{\pm\alpha}_{\tau^\kappa_{j+1}}-\Pi^{\pm \alpha}_{\tau^\kappa_j})S^1_{\tau^\kappa_{j+1}}
+ \kappa|\Pi^{\pm\alpha}_{\tau^\kappa_{j+1}}-\Pi^{\pm \alpha}_{\tau^\kappa_j}|
S^1_{\tau^\kappa_{j+1}}
+(\hat{\Pi}^{0,\pm \alpha,\kappa}_{j+1}-\hat{\Pi}^{0,\pm
 \alpha,\kappa}_{j})S^0_{\tau^\kappa_{j+1}} = 0.
\end{split}
\end{equation*}
Note that the second term of (\ref{price}) is to absorb the
transaction cost at time $0$.
The value $V^{\pm \alpha, \kappa}_t$ 
of our portfolio at time $t$
ignoring the clearance cost
is given by
\begin{equation*}
V^{\pm \alpha,\kappa}_t = \Pi^{\pm \alpha,\kappa}_t S^1_t + 
\Pi^{0,\pm \alpha,\kappa}_t S^0_t,
\end{equation*}
where we put $\Pi^{\pm \alpha,\kappa}_t = \Pi^{\pm \alpha}_{\tau^\kappa_j}$ and
$\Pi^{0,\pm \alpha,\kappa}_t = \hat{\Pi}^{0,\pm \alpha,\kappa}_j$ 
for $t \in [\tau^\kappa_j,\tau^\kappa_{j+1})$.
By construction we have that
\begin{equation*}
\Delta V^{\pm \alpha,\kappa}_j =
\Pi^{\pm \alpha}_{\tau^\kappa_j}\Delta S^1_j
+ \hat{\Pi}^{0,\pm \alpha,\kappa}_j \Delta S^0_j
- \kappa|\Delta \Pi^{\pm \alpha}_j|S^1_{\tau^\kappa_{j+1}},
\end{equation*}
where 
\begin{equation*}
\Delta V^{\pm \alpha,\kappa}_j = 
V^{\pm \alpha,\kappa}_{\tau^\kappa_{j+1}} - 
V^{\pm \alpha,\kappa}_{\tau^\kappa_j}, \ \ 
\Delta \Pi^{\pm \alpha}_j = 
\Pi^{\pm \alpha}_{\tau^\kappa_{j+1}}-
\Pi^{\pm \alpha}_{\tau^\kappa_{j}}, \ \ 
\Delta S^i_j = S^i_{\tau^\kappa_{j+1}}-S^i_{\tau^\kappa_j}
\end{equation*}
for $i = 0$ and $1$.
Further, putting 
\begin{equation*}
\tilde{V}^{\pm \alpha,\kappa} = V^{\pm \alpha,\kappa}/S^0, \ \ 
\Delta \tilde{V}^{\pm \alpha,\kappa}_j = 
\tilde{V}^{\pm \alpha,\kappa}_{\tau^\kappa_{j+1}} - 
\tilde{V}^{\pm \alpha,\kappa}_{\tau^\kappa_j}, \ \ 
\Delta \tilde{S}_j = \tilde{S}_{\tau^\kappa_{j+1}}-\tilde{S}_{\tau^\kappa_j},
\end{equation*}
we obtain that
\begin{equation*}
\Delta \tilde{V}^{\pm \alpha,\kappa}_j =
\Pi^{\pm \alpha}_{\tau^\kappa_j}\Delta \tilde{S}_j
- \kappa|\Delta \Pi^{\pm \alpha}_j|\tilde{S}_{\tau^\kappa_{j+1}},
\end{equation*}
so that
\begin{equation} \label{Vtil}
\tilde{V}^{\pm \alpha,\kappa}_t = P(S^1_0,R_0,\Sigma^{\pm \alpha}_0)/S^0_0
+  \int_0^t \Pi^{\pm \alpha,\kappa}_u \mathrm{d}\tilde{S}_u 
- \sum_{\tau^\kappa_{j+1} \leq t}
\kappa 
|\Pi^{\pm \alpha}_{\tau^\kappa_{j+1}}-\Pi^{\pm \alpha}_{\tau^\kappa_j}|
\tilde{S}_{\tau^\kappa_{j+1}}.
\end{equation}

Our main result concerns the limit distribution of the
continuous process $Z^{\pm \alpha,\kappa}$ defined as
\begin{equation*}
Z^{\pm \alpha,\kappa}_t
= \kappa^{-1}(P(S^1_t,R_t,\Sigma^{\pm \alpha}_t)/S^0_t - 
\tilde{V}^{\pm\alpha,\kappa}_t)
\end{equation*}
as $\kappa \to 0$, and in particular, we have a convergence
\begin{equation*}
V^{\pm \alpha}_t \to P(S^1_t,R_t,\Sigma^{\pm \alpha}_t)
\end{equation*}
in probability for $t \in [0,\tau)$.
Note that if $f$ is convex, then
\begin{equation*}
P(S^1_T,R_T,\Sigma^{\pm \alpha}_T)
=P(S^1_T,0,\Sigma^{\pm \alpha}_T) \geq f(S^1_T)
\end{equation*}
on the set $\{ \tau \geq T\}$.
If $f$ is concave, 
\begin{equation*}
P(S^1_{\tau},R_{\tau},\Sigma^{\pm\alpha}_{\tau}) = 
S^0_{\tau}f(\tilde{S}_{\tau}) = a S^1_{\tau} + bS^0_{\tau}, \ \ 
aS^1_T + bS^0_T \geq f(S^1_T) 
\end{equation*}
on the set $\{\tau \leq T\}$.
Therefore, in the asymptotic sense,
the self-financing strategy
$(\Pi^{\pm\alpha,\kappa}, \Pi^{0,\pm \alpha,\kappa})$ turns out to be
a super-hedging strategy for convex or concave payoffs
under the cumulative volatility bounds
$\langle \log(\tilde{S})\rangle_T \leq \hat{\Sigma}$ or
$\langle \log(\tilde{S})\rangle_T \geq \hat{\Sigma}$ respectively.
In other words, asymptotically this  serves as 
the conservative delta hedging strategy described
in Section~2.
The role of $\hat{\Sigma}$ is exactly the same.
Note that even outside of the specified set $\{ \tau \geq T\}$
or $\{ \tau \leq T\}$, the strategy is 
well-defined and the following
results remain valid.

Now we put a condition equivalent to 
\textit{ No Free Lunch with Vanishing Risk }
 (See Delbaen and
Schachermayer~\cite{DS1,DS2}),
one of ``no-arbitrage'' conditions which is natural to be
supposed in this financial context:
\begin{assu} \label{NA}
There exists an equivalent local martingale measure for $\tilde{S}$, that is,
an equivalent probability  measure under which $\tilde{S}$ is
 a local martingale.
\end{assu}
Denote by $D[0,\infty)$ the space of the cadlag functions on $[0,\infty)$
equipped with the Skorokhod topology.
Here we introduce the notion of ``stable convergence in $D[0,\sigma)$''
for a stopping time $\sigma$.
\begin{defn}
Let $\sigma$ be a stopping time and $\mathcal{G} \subset \mathcal{F}$ be
 a sub $\sigma$-field.
A family of continuous processes $Z^\kappa$, $\kappa > 0$ is said to converge 
$\mathcal{G}$-stably in $D[0,\sigma)$ to $Z$ as $\kappa \to 0 $ 
if there exists a sequence 
of stopping times $\sigma^m$ which converges to $\sigma$ a.s. as 
$m\to  \infty$ such that for any $m \in \mathbb{N}$,  
any bounded $\mathcal{G}$-measurable random 
 variable $G$ and any positive sequence $\kappa_n$ with $\kappa_n \to 0$
as $n \to \infty$, the $\mathbb{R} \times D[0,\infty)$-valued random sequence
$(G,Z^{\kappa_n}_{\cdot \wedge \sigma^m})$ 
converges to $(G,Z_{\cdot \wedge \sigma^m})$ in law as $n \to \infty$.
\end{defn}
For a family of random variables $X_\epsilon$, $\epsilon > 0$,
 we write $X_\epsilon = O_p(\epsilon)$ if $\epsilon^{-1}X_{\epsilon}$
is tight, that is, 
$\lim_{K\to \infty}\sup_{\epsilon \in (0,1]}
\mathbb{P}[\epsilon^{-1}|X_{\epsilon}| > K] = 0$.
Further, we write $X_\epsilon = o_p(\epsilon)$ if 
$\epsilon^{-1}X_{\epsilon} \to  0$ in probability as $\epsilon \to 0$.
\begin{thm}
Suppose Conditions \ref{scon} and \ref{NA} to hold. 
Let $W$ be a standard Brownian motion independent of
      $\mathcal{F}$, possibly defined on an extension 
of $(\Omega, \mathcal{F},\mathbb{P}, \{\mathcal{F}_t\})$. 
\begin{itemize}
\item If $f$ is convex, then 
$Z^{+\alpha,\kappa}$ converges $\mathcal{F}$-stably in 
      $D[0,\tau)$ to a time-changed Brownian motion
      $W_Q$ as $\kappa \to 0$,  where 
\begin{equation*}
Q = \frac{\left|\alpha + 2\right|^2}{6}
\int_0^{\cdot} |\tilde{S}_u\Gamma^{+\alpha}_u |^2\mathrm{d}
\langle \tilde{S} \rangle_u.
\end{equation*}

\item
If $f$ is concave, then 
$Z^{-\alpha,\kappa}$ converges $\mathcal{F}$-stably
in 
      $D[0,\tau)$ to a time-changed Brownian motion
      $W_Q$ as $\kappa \to 0$,  where  
\begin{equation*}
Q = \frac{\left|\alpha - 2\right|^2}{6}
\int_0^{\cdot} |\tilde{S}_u\Gamma^{-\alpha}_u |^2\mathrm{d}
\langle \tilde{S} \rangle_u.
\end{equation*}
\end{itemize}
\end{thm}
{\it Proof: }
We take a sequence of stopping times $\sigma^m$ so that
$\tilde{S}$ and $1/\tilde{S}$
are bounded by $m$ on $[0,\sigma^m]$ and
$\sigma^m \leq m$, 
$\langle \log(\tilde{S}) \rangle_{\sigma^m} \leq \hat{\Sigma}-1/m$
for each $m \in \mathbb{N}$ and 
$\sigma^m \to \tau$ a.s. as $m \to \infty$.
Notice that putting
\begin{equation*}
\Delta(S,\Sigma) = \frac{\partial P}{\partial S}(S,0,\Sigma) 
\end{equation*}
for $(S,\Sigma) \in \mathbb{R}_+^2$, 
we have
\begin{equation}\label{tilde}
\Pi^{\pm \alpha} = \Delta(\tilde{S},\Sigma^{\pm \alpha}), \ \ 
\Gamma^{\pm \alpha} = \frac{\partial \Delta}{\partial S}
(\tilde{S},\Sigma^{\pm \alpha}).
\end{equation}
Hence $\Pi^{\pm \alpha}$, $\Gamma^{\pm \alpha}$ and 
$1/\Gamma^{\pm \alpha}$  are bounded on $[0,\sigma^m]$.
By It$\hat{\text{o}}$'s formula,
\begin{equation*}
\begin{split}
P(S^1_t, R_t, \Sigma^{\pm \alpha}_t)
=& P(S^1_0, R_0, \Sigma^{\pm \alpha}_0) \\
&+  \int_0^t
\Pi^{\pm \alpha}_u \mathrm{d}S^1_u + 
\int_0^t
\Pi^{0,\pm \alpha}_u \mathrm{d}S^0_u  \mp \frac{1}{\alpha}
\int_0^t S^0_u\Gamma^{\pm \alpha}_u
\mathrm{d}\langle \tilde{S} \rangle_u
\end{split}
\end{equation*}
for $t \in [0,\tau)$.
Hence, again by It$\hat{\text{o}}$'s formula, 
\begin{equation*}
\tilde{P}^{\pm \alpha}_t = 
\tilde{P}^{\pm \alpha}_0 + \int_0^t \Pi^{\pm \alpha}_u \mathrm{d}\tilde{S}_u
\mp \frac{1}{\alpha}
\int_0^t \Gamma^{\pm \alpha}_u
\mathrm{d}\langle \tilde{S} \rangle_u,
\end{equation*}
where $\tilde{P}^{\pm \alpha} = P(S^1,R,\Sigma^{\pm \alpha})/S^0$. 
Using (\ref{dates}) and (\ref{Vtil}),
 we have that
\begin{equation} \label{error}
\begin{split}
\tilde{P}^{\pm \alpha}_t
- \tilde{V}^{\pm \alpha,\kappa}_t
=&  \int_0^t(\Pi^{\pm \alpha}_u -
 \Pi^{\pm \alpha,\kappa}_u)\mathrm{d} \tilde{S}_u 
\\ & +  \alpha \kappa^2
\sum_{\tau^\kappa_{j+1} \leq t } \tilde{S}_{\tau^\kappa_{j+1}}\tilde{S}_{\tau^\kappa_j}
|\Gamma^{\pm \alpha }_{\tau^\kappa_j}|
\mp
\frac{1}{\alpha}
\int_0^t \frac{ \mathrm{d}\langle \Pi^{\pm \alpha} \rangle_u}
{\Gamma^{\pm \alpha}_u}.
\end{split}
\end{equation}
Since stable convergence and in particular, convergence in probability are
 stable against equivalent changes of
measures, 
we can freely choose an equivalent measure to estimate terms.
First take an equivalent martingale measure for $\tilde{S}$.
Then,
\begin{equation*}
\begin{split}
&\mathbb{E}\left[ \sup_{0 \leq t \leq \sigma^m}\left|
\sum_{\tau^\kappa_{j+1} \leq t } \tilde{S}_{\tau^\kappa_j}
|\Gamma^{\pm \alpha }_{\tau^\kappa_j}|
(\tilde{S}_{\tau^\kappa_{j+1}}-\tilde{S}_{\tau^\kappa_{j}})
\right|^2\right] \\
&\leq C_m + 
\mathbb{E}\left[ \sup_{0 \leq t \leq \sigma^m}\left|
\sum_{j=0}^{\infty} \tilde{S}_{\tau^\kappa_j}
|\Gamma^{\pm \alpha }_{\tau^\kappa_j}|
(\tilde{S}_{\tau^\kappa_{j+1} \wedge t}-\tilde{S}_{\tau^\kappa_{j} \wedge t})
\right|^2\right]\\
&\leq  C_m + 
\mathbb{E}\left[ \
\sum_{j=0}^{\infty} \tilde{S}_{\tau^\kappa_j}^2
|\Gamma^{\pm \alpha }_{\tau^\kappa_j}|^2
(\langle \tilde{S} \rangle_{\tau^\kappa_{j+1} \wedge \sigma^m}-
\langle \tilde{S} \rangle_{\tau^\kappa_{j} \wedge \sigma^m}) \right]
\end{split}
\end{equation*}
by Doob's inequality, 
where $C_m$ is a constant.
Therefore, using (\ref{dates}) again, 
we have that the sum of the last two terms  of (\ref{error}) is equal to
\begin{equation*}
\pm \frac{1}{\alpha}
\left\{
\sum_{j=0}^{\infty} \frac{1}{\Gamma^{\pm \alpha}_{\tau^\kappa_j}}
|\Pi^{\pm \alpha}_{t \wedge \tau^\kappa_{j+1}} - \Pi^{\pm \alpha}_{t \wedge
\tau^\kappa_j} | ^2 -
\int_0^t \frac{ \mathrm{d}\langle \Pi^{\pm \alpha} \rangle_u}
{\Gamma^{\pm \alpha}_u}
\right\} + O_p(\kappa^2).
\end{equation*}
Here we have used $|\Gamma^{\pm \alpha}| = \pm \Gamma^{\pm \alpha}$.
Since
\begin{equation*}
|\Pi^{\pm \alpha}_{t} - 
\Pi^{\pm \alpha}_{ s}|^2
=
2 \int_{s}^{t}
(\Pi^{\pm \alpha}_u- \Pi^{\pm \alpha}_s) \mathrm{d}\Pi^{\pm \alpha}_u
+ \langle \Pi^{\pm \alpha} \rangle_{t}
- \langle \Pi^{\pm \alpha} \rangle_{s},
\end{equation*}
we obtain that
\begin{equation} \label{error2}
\begin{split}
Z^{\pm \alpha,\kappa}_t =& \kappa^{-1}\left\{1 \pm \frac{2}{\alpha}\right\}
\int_0^t (\Pi^{\pm \alpha}_u-\Pi^{\pm
 \alpha,\kappa}_u)\mathrm{d}\tilde{S}_u \\
& \pm \frac{1}{\alpha \kappa} \left\{
\int_0^t \frac{ \mathrm{d}\langle \Pi^{\pm \alpha} \rangle_u}
{\Gamma^{\pm \alpha,\kappa}_u} -
\int_0^t \frac{ \mathrm{d}\langle \Pi^{\pm \alpha} \rangle_u}
{\Gamma^{\pm \alpha}_u}
\right\} \\
&\pm \frac{2}{\alpha \kappa}\left\{
\int_0^t \frac{\Pi^{\pm \alpha}_u - \Pi^{\pm \alpha,\kappa}_u}{
\Gamma^{\pm \alpha,\kappa}_u}\mathrm{d}\Pi^{\pm
 \alpha}_u
-
 \int_0^t (\Pi^{\pm \alpha}_u-\Pi^{\pm \alpha,\kappa}_u)\mathrm{d}\tilde{S}_u
\right\} + O_p(\kappa),
 \end{split}
\end{equation}
where 
$\Gamma^{\pm \alpha,\kappa}_t = \Gamma^{\pm \alpha}_{\tau^\kappa_j}$
for $t \in [\tau^\kappa_j, \tau^\kappa_{j+1})$.
Next for each of $+ \alpha$ and  $-\alpha$,
take an equivalent measure under which
 $\Pi^{\pm \alpha}_{\cdot \wedge \sigma^m}$ is a local martingale.
Then, by Lemma~\ref{key} below, we have that
\begin{equation*}
\int_0^t \frac{ \mathrm{d}\langle \Pi^{\pm \alpha} \rangle_u}
{\Gamma^{\pm \alpha}_u}
= 
\int_0^t \frac{ \mathrm{d}\langle \Pi^{\pm \alpha} \rangle_u}
{\Gamma^{\pm \alpha,\kappa}_u}
+o_p(\kappa)
\end{equation*}
and that
\begin{equation*}
\int_0^t \frac{\Pi^{\pm \alpha}_u - \Pi^{\pm \alpha,\kappa}_u}{
\Gamma^{\pm \alpha,\kappa}_u}\mathrm{d}\Pi^{\pm
 \alpha}_u
= 
 \int_0^t (\Pi^{\pm \alpha}_u-\Pi^{\pm \alpha,\kappa}_u)\mathrm{d}\tilde{S}_u
+ o_p(\kappa).
\end{equation*}
The result then follows from Theorem~2.6 of Fukasawa~\cite{aap}.
\hfill////

\begin{lem} \label{key}
Suppose Conditions~\ref{scon} and \ref{NA} to hold.
Let $\sigma^m$ be a bounded stopping time such that
$\tilde{S}$ and $1/\tilde{S}$
are bounded by $m$ on $[0,\sigma^m]$ and
$\langle \log(\tilde{S}) \rangle_{\sigma^m} \leq \hat{\Sigma}-1/m$.
Denote by $\mathbb{E}^{\pm \alpha}$ the expectation operators
of equivalent measures under which 
$\Pi^{\pm \alpha }_{\cdot \wedge \sigma^m}$
are local martingales respectively.
Let $U$, $V$ be twice continuously differentiable functions on 
$(0,\infty)^2$.
Set
\begin{equation*}
U^{\pm \alpha}_t = U(\tilde{S}_t,\Sigma^{\pm \alpha}_t), \ \ 
V^{\pm \alpha}_t = V(\tilde{S}_t,\Sigma^{\pm \alpha}_t)
\end{equation*}
and $U^{\pm \alpha,\kappa}_t = U^{\pm \alpha}_{\tau^\kappa_j}$ 
for $t \in [\tau^\kappa_j , \tau^\kappa_{j+1})$.
Then for any $k \in \mathbb{N}$,
\begin{equation} \label{finite}
\sup_{ \kappa \in (0,1]}
\kappa^{-2k}
\mathbb{E}^{\pm \alpha}
\left[
\int_0^{\sigma^m} |U^{\pm \alpha}_s - U^{\pm \alpha}_s|^{2k}
\mathrm{d}\langle \Pi^{\pm \alpha } \rangle_s
\right]  < \infty
\end{equation}
and
\begin{equation} \label{conv0}
\lim_{\kappa \to 0} \kappa^{-2}
\mathbb{E}^{\pm \alpha}
\left[
\sup_{t \geq 0} \left|
\int_0^{t \wedge \sigma^m} (U^{\pm \alpha}_s - U^{\pm \alpha,
\kappa}_s)V^{\pm \alpha}_s
\mathrm{d}\langle \Pi^{\pm \alpha } \rangle_s
\right|^2
\right] = 0.
\end{equation}
\end{lem}
{\it Proof: }
We omit $\pm \alpha$ since everything is the same for the both cases.
We use $C$ as a generic positive constant which does not depend on $n$.
To prove this lemma we can 
suppose without loss of generality that
\begin{equation*}
\langle \Pi \rangle_t = \langle \Pi \rangle_{t \wedge 
\sigma^m} + (t - \sigma^m)_+, \ \ 
\tilde{S}_t = \tilde{S}_{t \wedge \sigma^m}, \ \ 
\Gamma_t = \Gamma_{t \wedge \sigma^m}
\end{equation*}
for all $t \geq0$.
By the Burkholder-Davis-Gundy, and Doob's inequalities, we have
\begin{equation*}
C^{-1} \kappa^{2k} \leq 
\mathbb{E}[|\langle \Pi \rangle_{\tau^\kappa_{j+1}} 
- \langle \Pi \rangle_{\tau^\kappa_j}|^k | \mathcal{F}_{\tau^\kappa_j}]
\leq C \kappa^{2k}
\end{equation*}
uniformly in $j = 0, 1, \dots, N^\kappa_t$ for any $t \geq 0$, where 
\begin{equation*}
N^\kappa_t = \max\{j \geq 0; \tau^\kappa_j \leq t \}.
\end{equation*}
Moreover,
\begin{equation*}
\kappa^2 \mathbb{E}[N^\kappa_t] 
\leq \mathbb{E}\left[
\sum_{j=0}^{N^\kappa_t} \frac{|\Pi_{\tau^\kappa_{j+1}} - 
\Pi_{\tau^\kappa_j}|^2}{ \alpha^2 \tilde{S}_{\tau^\kappa_j}^2 \Gamma_{\tau^\kappa_j}^2}
\right]  \leq Ct + 
C \mathbb{E}[\langle \Pi \rangle_{\sigma^m}],
\end{equation*}
so that
\begin{equation} \label{quad}
\begin{split}
\mathbb{E}\left[
\sum_{j=0}^{\infty} |\langle \Pi \rangle_{t \wedge \tau^\kappa_{j+1}}
- \langle \Pi \rangle_{t \wedge \tau^\kappa_{j}}|^k
\right]
= &
\mathbb{E}\left[
\sum_{j=0}^{\infty} \mathbb{E}[|\langle \Pi \rangle_{t \wedge \tau^\kappa_{j+1}}
- \langle \Pi \rangle_{t \wedge \tau^\kappa_{j}}|^k | \mathcal{F}_{t \wedge 
\tau^\kappa_j}]
\right] \\
\leq &
\mathbb{E}\left[
\sum_{j=0}^{\infty}\mathbb{E}[|\langle \Pi \rangle_{\tau^\kappa_{j+1}}
- \langle \Pi \rangle_{\tau^\kappa_{j}}|^k | \mathcal{F}_{
\tau^\kappa_j}] 1_{\{j \leq N^\kappa_{t}\}}
\right] \\
\leq  & C\kappa^{2k}(1 + \mathbb{E}[N^\kappa_t]) = O(\kappa^{2(k-1)})
\end{split}
\end{equation}
for any $k \in \mathbb{N}$.
Hence the same argument as in the proof of Lemma~2.9 of Fukasawa~\cite{aap}
is applicable to obtain (\ref{finite}) in cases where there exists 
a continuous bounded process $Y$ such that
\begin{equation*}
\mathrm{d}U_t = Y_t \mathrm{d}\Pi_t
\end{equation*}
on $[0,\sigma^m]$.
In general, there exist continuous bounded processes $Y^i$
such that
\begin{equation*}
\mathrm{d}U_t = Y^1_t \mathrm{d}\Pi_t + Y^2_t \mathrm{d}\langle \Pi
 \rangle_t, \ \ 
\mathrm{d}V_t = Y^3_t \mathrm{d}\Pi_t + Y^4_t \mathrm{d}\langle \Pi
 \rangle_t
\end{equation*}
on $[0,\sigma^m]$.
To obtain (\ref{finite}) in the general case, observe that
\begin{equation*} 
\lim_{ \kappa \to 0}
\kappa^{-2k}
\mathbb{E}
\left[\sum_{j=0}^{\infty}
\int_{\tau^\kappa_j \wedge \sigma^m}^{\tau^\kappa_{j+1}\wedge \sigma^m} \left|
\int_{\tau^\kappa_j \wedge \sigma^m}^s Y^2_u \mathrm{d}\langle \Pi \rangle_u
\right|^{2k}
\mathrm{d}\langle \Pi^{\pm \alpha } \rangle_s
\right]  = 0.
\end{equation*}
This is because $Y^i$ are bounded and we already have (\ref{quad}).
By the same reason, we can suppose $Y^4=0$
to prove (\ref{conv0}). Further,
we can replace $V = V^{\pm \alpha}$ with $V^\kappa$ by (\ref{finite}) 
with the aid of the
Cauchy-Schwarz inequality,
where $V^\kappa_t = V_{\tau^\kappa_j}$ for $t \in [\tau^\kappa_j, \tau^\kappa_{j+1})$.
It suffices then to show that
\begin{equation*}
\lim_{\kappa \to 0} \kappa^{-2}
\mathbb{E}
\left[ \sup_{t \geq 0} \left|
\sum_{j=0}^{\infty}
\int_{\tau^\kappa_j \wedge \sigma^m}^{\tau^\kappa_{j+1}\wedge \sigma^m} 
\int_{\tau^\kappa_j \wedge \sigma^m}^t
(Y^3_s-Y^3_{\tau^\kappa_j})\mathrm{d} \Pi_s
V_{\tau^\kappa_j} \mathrm{d}\langle \Pi \rangle_t
\right|^2
\right] = 0
\end{equation*}
and that
\begin{equation*}
\lim_{\kappa \to 0} \kappa^{-2}
\mathbb{E}
\left[ \sup_{t \geq 0} \left|
\sum_{j=0}^{N^\kappa_{t \wedge \sigma^m}}
\int_{\tau^\kappa_j}^{\tau^\kappa_{j+1}} 
(\Pi_t-\Pi_{\tau^\kappa_j})Y^3_{\tau^\kappa_j}
V_{\tau^\kappa_j} \mathrm{d}\langle \Pi \rangle_t
\right|^2
\right] = 0.
\end{equation*}
The first one follows again from the same argument as in the
proof of Lemma~2.9 of
Fukasawa~\cite{aap}.
To show the second, notice that by the definition (\ref{dates}) and 
the martingale property of $\Pi$,
\begin{equation*}
 \mathbb{E}\left[
\int_{\tau^\kappa_j}^{\tau^\kappa_{j+1}} (\Pi_t - \Pi_{\tau^\kappa_j})\mathrm{d}
\langle \Pi \rangle_t
\Bigg| \mathcal{F}_{\tau^\kappa_j}
\right] = \frac{1}{3}
\mathbb{E}[(\Pi_{\tau^\kappa_{j+1}}-\Pi_{\tau^\kappa_j})^3|\mathcal{F}_{\tau^\kappa_j}]
=0.
\end{equation*}
Therefore, we obtain the result by observing that
\begin{equation*}
\begin{split}
&\mathbb{E}
\left[ 
\sum_{j=0}^{ \infty }
\left| \int_{\tau^\kappa_j \wedge \sigma^m}^{\tau^\kappa_{j+1}\wedge \sigma^m} 
(\Pi_t-\Pi_{\tau^\kappa_j})Y^3_{\tau^\kappa_j}
V_{\tau^\kappa_j} \mathrm{d}\langle \Pi \rangle_t
\right|^2
\right]\\
& \leq
C \mathbb{E}
\left[ 
\sum_{j=0}^{\infty}
|\langle \Pi \rangle_{\tau^\kappa_{j+1} \wedge \sigma^m} - 
\langle \Pi \rangle_{\tau^\kappa_j \wedge \sigma^m}|^{3/2}
\left\{
\int_{\tau^\kappa_{j}\wedge \sigma^m}^{\tau^\kappa_{j+1} \wedge \sigma^m}
|\Pi_t-\Pi_{\tau^\kappa_j}|^4 \mathrm{d}
\langle \Pi \rangle_t
\right\}^{1/2}
\right]\\
&\leq C \left\{ \mathbb{E}\left[
\sum_{j=0}^{ \infty}
|\langle \Pi \rangle_{\tau^\kappa_{j+1} \wedge \sigma^m} 
- \langle \Pi \rangle_{\tau^\kappa_j \wedge \sigma^m}|^3
\right] \right\}^{1/2} \left\{ \mathbb{E}\left[
\int_0^{\sigma^m} |\Pi_t-\Pi^\kappa_t|^4 \mathrm{d} \langle \Pi \rangle_t
\right]\right\}^{1/2} = o(\kappa^2)
\end{split} 
\end{equation*}
because of (\ref{quad}), where $\Pi^\kappa = \Pi^{\pm \alpha,\kappa}$. 
\hfill////

\section{Mean squared error}
Here we 
study the mean squared error of
the discrete hedging strategy under linear transaction costs described in the
last section.
We prove the following convergence,
that is formally indicated by the results of the last section.
\begin{thm} \label{MSE}
Suppose Conditions~\ref{scon} and $\ref{NA}$ to hold. 
Denote by $\mathbb{E}$ the expectation with respect to 
the equivalent local martingale measure for
 $\tilde{S}$. Then,
there exists a sequence of stopping times $\sigma^m$ such that
$\sigma^m \to \tau$ a.s. as $m \to \infty$ and
\begin{equation*}
\lim_{\kappa \to 0} \mathbb{E}[
| Z^{\pm \alpha,\kappa }_{t \wedge \sigma^m} |^2]   = \frac{|\alpha \pm 2|^2}{6}
\mathbb{E}\left[
\int_0^{t \wedge \sigma^m} |\tilde{S}_u \Gamma^{\pm \alpha}_u|^2
 \mathrm{d}\langle \tilde{S} \rangle_u
\right]
\end{equation*}
for any $t \geq 0$, 
where as before, $\pm \alpha$ should be understood as $+ \alpha$ or 
$- \alpha$ if $f$ is convex or concave respectively.
\end{thm}
{\it Proof: }
For given a continuous semimartingale $X$ with decomposition
\begin{equation*}
X_t = X_0 + M_t+ \int_0^t V_t \mathrm{d}\langle M \rangle_t,
\end{equation*}
where $M$ is a continuous local martingale and $V$ is 
a locally bounded adapted process, 
define a continuous process $e(X)$ as
\begin{equation*}
e(X)_t = \exp\left\{
\int_0^t V_t \mathrm{d}M_t- \frac{1}{2} \int_0^t
V_t^2 \mathrm{d}\langle M \rangle_t
\right\}.
\end{equation*}
Take a sequence of stopping times $\sigma^m$ so that
$\tilde{S}$, $1/\tilde{S}$ and $1/e(\Pi)$ are bounded by $m$
on $[0,\sigma^m]$ and
$\sigma^m \leq m$, $\langle \log(\tilde{S})\rangle_{\sigma^m} \leq
\hat{\Sigma} -1/m$ 
for each $m$, and $\sigma^m \to \tau$ a.s.~as $m \to \infty$.
We may start from (\ref{error2}),
where the $O_p(\kappa)$ term is negligible in $L^2$.
Denote by $R^\kappa$ the sum of the second and third terms of (\ref{error2}).
By Lemma~\ref{key} and the Girsanov-Maruyama theorem, 
we have that
\begin{equation*}
\mathbb{E}[|R^\kappa_{t \wedge \sigma^m}|^2] \leq m 
\mathbb{E}[e(\Pi)_{t \wedge \sigma^m}|R^\kappa_{t \wedge \sigma^m}|^2]
\to 0
\end{equation*}
as $\kappa \to 0$. Now it remains to show
\begin{equation*}
\lim_{\kappa \to \infty} \kappa^{-2}
\mathbb{E}\left[
\int_0^{t \wedge \sigma^m}
(\Pi^{\pm \alpha}_u- \Pi^{\pm \alpha,\kappa}_u)^2 \mathrm{d}
\langle \tilde{S} \rangle_u
\right] 
= \frac{\alpha^2}{6}
\mathbb{E}\left[
\int_0^{t \wedge \sigma^m} |\tilde{S}_u \Gamma^{\pm \alpha}_u|^2
 \mathrm{d}\langle \tilde{S} \rangle_u
\right].
\end{equation*}
By Lemma~\ref{key} again, we have that
\begin{equation*}
\begin{split}
\kappa^{-2}\mathbb{E}&\left[
\int_0^{t \wedge \sigma^m}
(\Pi^{\pm \alpha}_u- \Pi^{\pm \alpha,\kappa}_u)^2 \left|
\frac{1}{|\Gamma^{\pm \alpha}_u|^2} - \frac{1}{|\Gamma^{\pm \alpha,\kappa}_u|^2}
\right| \mathrm{d}
\langle \tilde{\Pi}^{\pm \alpha} \rangle_u
\right] \\
\leq m \kappa^{-2} & \left|\mathbb{E}\left[e(\Pi)_{t \wedge \sigma^m}
\int_0^{t \wedge \sigma^m}
(\Pi^{\pm \alpha}_u- \Pi^{\pm \alpha,\kappa}_u)^4 \mathrm{d}
\langle \tilde{\Pi}^{\pm \alpha} \rangle_u
\right] 
 \right|^{1/2}  \\ & \times
\left|\mathbb{E}\left[e(\Pi)_{t \wedge \sigma^m}
\int_0^{t \wedge \sigma^m}
\left|
\frac{1}{|\Gamma^{\pm \alpha}_u|^2} - \frac{1}{|\Gamma^{\pm \alpha,\kappa}_u|^2}
\right|^2
\mathrm{d}
\langle \tilde{\Pi}^{\pm \alpha} \rangle_u
\right]
 \right|^{1/2} = O(\kappa).
\end{split}
\end{equation*}
By It$\hat{\text{o}}$'s formula, we have that
\begin{equation*}
\left(\Pi^{\pm \alpha}_{\tau^\kappa_{j+1}} - \Pi^{\pm \alpha}_{\tau^\kappa_j}
\right)^4=
4 \int_{\tau^\kappa_j}^{\tau^\kappa_{j+1}}
\left(\Pi^{\pm \alpha}_s - \Pi^{\pm \alpha}_{\tau^\kappa_j}
\right)^3 \mathrm{d}\Pi^{\pm \alpha}_s + 6
\int_{\tau^\kappa_j}^{\tau^\kappa_{j+1}}
\left(\Pi^{\pm \alpha}_s - \Pi^{\pm \alpha}_{\tau^\kappa_j}
\right)^2 \mathrm{d}\langle \Pi^{\pm \alpha} \rangle_s,
\end{equation*}
so that putting $N^\kappa_s = \max\{j \geq 0; \tau^\kappa_j \leq s\}$,
\begin{equation*}
\begin{split}
&\lim_{\kappa \to \infty} \kappa^{-2}
\mathbb{E}\left[
\int_0^{t \wedge \sigma^m}
(\Pi^{\pm \alpha}_u- \Pi^{\pm \alpha,\kappa}_u)^2 \mathrm{d}
\langle \tilde{S} \rangle_u
\right] \\
&=
\lim_{\kappa \to \infty} \kappa^{-2}
\mathbb{E}\left[
\sum_{j=0}^{N^\kappa_{t \wedge \sigma^m}}
\int_{\tau^\kappa_j}^{ \tau^\kappa_{j+1}}
\frac{(\Pi^{\pm \alpha}_u- \Pi^{\pm \alpha}_{\tau^\kappa_j})^2}
{|\Gamma^{\pm \alpha}_{\tau^\kappa_j}|^2} \mathrm{d}
\langle \Pi^{\pm \alpha} \rangle_u
\right] \\
&= \frac{1}{6}
\lim_{\kappa \to 0} \kappa^{-2}
\mathbb{E}\left[
\sum_{j=0}^{N^\kappa_{t \wedge \sigma^m}}
\frac{\left|
\Pi^{\pm \alpha}_{\tau^\kappa_{j+1}} - \Pi^{\pm \alpha}_{\tau^\kappa_j}
\right|^4}{|\Gamma^{\pm \alpha}_{\tau^\kappa_j}|^2}
\right]\\
&= \frac{\alpha^2}{6}
\lim_{\kappa \to 0}
\mathbb{E}\left[
\sum_{j=0}^{N^\kappa_{t \wedge \sigma^m}}
|\tilde{S}_{\tau^\kappa_j}|^2
\left|
\Pi^{\pm \alpha}_{\tau^\kappa_{j+1}} - \Pi^{\pm \alpha}_{\tau^\kappa_j}
\right|^2
\right].
\end{split}
\end{equation*}
By Lemma~\ref{key} again, we obtain that
\begin{equation*}
\begin{split}
&\mathbb{E}\left[\left|
\sum_{j=0}^{\infty}
|\tilde{S}_{\tau^\kappa_j}|^2
\left|
\Pi^{\pm \alpha}_{\tau^\kappa_{j+1}\wedge t \wedge \sigma^m} 
- \Pi^{\pm \alpha}_{\tau^\kappa_j
 \wedge t \wedge \sigma^m}
\right|^2
-  \int_0^{t \wedge \sigma^m}
|\tilde{S}^\kappa_u|^2 \mathrm{d}\langle \Pi^{\pm \alpha}\rangle_u
\right|^2\right]\\
&\leq 2m
\mathbb{E}\left[
e(\Pi)_{t \wedge \sigma^m}
\int_0^{t \wedge \sigma^m}
\left|
\Pi^{\pm \alpha}_u 
- \Pi^{\pm \alpha,\kappa}_u
\right|^2 |\tilde{S}^\kappa_u|^2
\mathrm{d}\langle \Pi^{\pm \alpha} \rangle_u
\right] = O(\kappa^2),
\end{split}
\end{equation*}
where $\tilde{S}^\kappa_t = \tilde{S}_{\tau^\kappa_j}$ for
$t \in [\tau^\kappa_j, \tau^\kappa_{j+1})$.
The rest is obvious. \hfill////

Unfortunately, we cannot claim in general that
\begin{equation*}
\lim_{\kappa \to 0} \mathbb{E}[
| Z^{\pm \alpha,\kappa }_{\tau} |^2]   = \frac{|\alpha \pm 2|^2}{6}
\mathbb{E}\left[
\int_0^{\tau} |\tilde{S}_u \Gamma^{\pm \alpha}_u|^2
 \mathrm{d}\langle \tilde{S} \rangle_u
\right].
\end{equation*}
In particular, even under the Black-Scholes model with volatility
$\sigma = \sqrt{\hat{\Sigma}/T}$,
we do not have in general that
\begin{equation*}
\lim_{\kappa \to 0} \kappa^{-2}\mathbb{E}[
| f(S_T^1)-V^{\pm \alpha}_T |^2]   = \frac{|\alpha \pm 2|^2}{6}
\mathbb{E}\left[
\int_0^T |\tilde{S}_u \Gamma^{\pm \alpha}_u|^2
 \mathrm{d}\langle \tilde{S} \rangle_u
\right].
\end{equation*}
This is mainly because we do not have enough tools to
 estimate the integrability of random variables depending on
the stopping times  $\tau^\kappa_j$
under reasonable assumptions on the regularity of $f$. 
Nevertheless,
it is practically irrelevant to
dynamically rebalance portfolios up to the 
last moment $T$.
It is common to avoid frequent rebalancing near the maturity due to the
high sensitivity of strategies with respect to price movement.
We are therefore satisfied with such a localized result as
 Theorem~\ref{MSE}.

\section{Leland's strategy and the choice of $\alpha$}
Our rebalancing dates (\ref{dates}) are not deterministic, 
while only deterministic dates have been treated in the 
preceding studies of Leland's strategy.
Therefore, our convergence results
are new even under the Black-Scholes model
that is a special case of our framework.
Here we compare the asymptotic variances of hedging error 
associated with our strategy and Leland's original one
and make a remark on the choice of $\alpha$ in (\ref{dates}).
Consider the Black-Scholes model with volatility $\sigma$ for 
the risk-neutral dynamics of $\tilde{S}$ and
let $f$ be
a convex payoff function.
Let $T=1$ for brevity.
Leland's original strategy uses the equidistant partition
$j/n$, $j=0,1,2,\dots,n$ as rebalancing date
and employs the Black-Scholes pricing and hedging strategy 
with enlarged volatility $\check{\sigma}$ defined as
\begin{equation*}
\check{\sigma}^2 = \sigma^2 + \sigma n^{1/2}\kappa \sqrt{\frac{8}{\pi}}.
\end{equation*}
In the case $\kappa = \kappa_n=\kappa_0n^{-1/2}$ with constant $\kappa_0>0$, 
it is known that
\begin{equation*}
\lim_{n \to \infty} \kappa_n^{-2}
\mathbb{E}[|f(S^1_1)-V^n_1|^2]
 = \beta \left(\frac{\sigma}{\kappa_0}\right) 
\mathbb{E}\left[
\int_0^1 |\tilde{S}_u \check{\Gamma}_u|^2 \mathrm{d}\langle \tilde{S} \rangle_u
\right],
\end{equation*}
where $V^n_t$ is the portfolio value at time $1$ of the strategy,
\begin{equation*}
\beta(x) = \frac{x^2}{2}+\sqrt{\frac{2}{\pi}}x
+ 1-\frac{2}{\pi}
\end{equation*}
 and
$\check{\Gamma}$ is defined as
\begin{equation*}
\check{\Gamma}_t = \frac{\partial^2 P}{\partial S^2}
(\tilde{S}_t,0,\check{\sigma}^2(1-t)).
\end{equation*}
See Gamys and Kabanov~\cite{GamysKabanov}.
Further, denoting by 
$\check{V}_t = P(\tilde{S}_t,0,\check{\sigma}^2{(1-t)})$, 
it is known that $\kappa_n^{-1}(\check{V}-V^n)$ converges stably in
$D[0,1]$ to a time changed Brownian motion $W_Q$ with
\begin{equation*}
Q= \beta \left(\frac{\sigma}{\kappa_0}\right) 
\int_0^{\cdot} |\tilde{S}_u \check{\Gamma}_u|^2 \mathrm{d}\langle
 \tilde{S} \rangle_u.
\end{equation*}
See Denis and Kabanov~\cite{DenisKabanov}.
Letting $\hat{\Sigma} = \sigma^2$ and 
\begin{equation}\label{alphakappa}
\alpha = \frac{\sigma}{\kappa_0}\sqrt{\frac{\pi}{2}},
\end{equation}
our strategy $(\Pi^{+ \alpha},\Pi^{0,+\alpha})$ 
has the same initial value as Leland's and
we have $\Gamma^{+\alpha} = \check{\Gamma}$.
The coefficient of the asymptotic variance is
\begin{equation*}
\frac{|\alpha+2|^2}{6} = \hat{\beta}\left(
\frac{\sigma}{\kappa_0}
\right), \ \ 
\hat{\beta}(x) = \frac{\pi}{12}x^2 + \frac{\sqrt{2\pi}}{3}x + \frac{2}{3}.
\end{equation*}
Since $1/2 > \pi/12$,
 if $\sigma/\kappa_0$, or
equivalently, $\alpha$ is sufficiently large, 
our strategy results in a smaller hedging error.
For example,  $\alpha \geq 2$ is enough to get this superiority.
To set $\alpha \geq 2$ is equivalent to set 
$ \check{\sigma}^2 \leq  2\sigma^2$,
which is indeed a realistic region of enlarged volatility.
We may examine more concrete values.
An approximate number of daily rebalancing $n = 250$ and
$1\%$ transaction costs $\kappa = 0.01$ 
yield $\kappa_0 \approx 0.15$. 
Then, estimates $\sigma = 0.20$ and $\sigma = 0.25$ 
give $\alpha \approx 1.67 $ and $\alpha \approx 2.01$ respectively. 
For both cases, we have the superiority:
\begin{equation*}
\begin{split}
&\beta\left( \frac{0.20}{0.15}\right) \approx 2.32 > 2.25 \approx 
\hat{\beta}\left( \frac{0.20}{0.15}\right),\\
&\beta\left( \frac{0.25}{0.15}\right) \approx 3.08 > 2.79 \approx 
\hat{\beta}\left( \frac{0.25}{0.15}\right).
\end{split}
\end{equation*}
As shown in Fukasawa~\cite{aap}, hitting times such as (\ref{dates})
always results in a smaller discretization error of stochastic integrals
than deterministic partitions.
Roughly speaking, this is because 
the asymptotic variance of the discretization error
is determined by the skewness and kurtosis of 
each increment of the integrand.
Hitting times produce the increments with
 unit kurtosis and zero skewness.
On the other hand, we are now dealing with not only the discretization error
 but also the linear transaction costs.
Due to the linear structure,
the first absolute moment of each increment also affects 
the asymptotic variance, while this is not clear from our proof.
This is the reason why the equidistant partition can be superior when
 $\alpha$ is small. 
Recall that in our framework, $\alpha$ can be chosen arbitrarily.
As $\alpha$ increases, our enlarging or shrinking volatility becomes
 modest and the price of the option decreases.
This occurs  at the expense of a larger asymptotic variance of 
the super-hedging error, which is clearly seen in our main theorems.
The optimal choice of $\alpha$ therefore depends on 
the risk preference of hedgers.

Finally we remark that the the number of rebalancing required by
our hitting time strategy (\ref{dates}) is not deterministic but 
asymptotically so and smaller than 
Leland's deterministic one subject to the common
initial value of strategy.
In fact for any $t \in [0,1)$,
\begin{equation*}
\begin{split}
\kappa^{-2} N^\kappa_t  &= \sum_{j=0}^{N^\kappa_t}
\frac{\left|\Pi^{\pm \alpha}_{\tau^\kappa_{j+1}} - \Pi^{\pm
\alpha}_{\tau^n_j}\right|^2}
{\alpha^2 \tilde{S}_{\tau^\kappa_j}^2 |\Gamma^{\pm
\alpha}_{\tau^\kappa_j}|^2} + o_p(1) \\
& \to \frac{1}{\alpha^2} \int_0^t 
\frac{1}{\tilde{S}_u^2 |\Gamma^{\pm\alpha}_u|^2}
\mathrm{d} \langle \Pi^{\pm \alpha} \rangle_u = \frac{1}{\alpha^2}
 \langle \log( \tilde{S}) \rangle_t
\end{split}
\end{equation*}
in probability as $\kappa \to 0$.
Under the Black-Scholes model we are now assuming,
 we have that
\begin{equation*}
n N^{\kappa_n}_t \to \frac{2}{\pi}t < t, \ \ \kappa_n = \kappa_0 n^{-1/2}
\end{equation*}
in probability as $n \to \infty$
with the specification (\ref{alphakappa}).
Recall that $n$ is the number of the equi-distant 
partition of $[0,1]$ for Leland's strategy.
Therefore our strategy (\ref{dates}) requires less frequent rebalancing,
which is another advantage.

\section{Conclusion}
We  proposed an explicit discrete hedging strategy for European
options with convex or concave payoff under a continuous semi-martingale
model with uncertainty and small transaction costs,
and determined the limit
distribution of the associated hedging error as the coefficient $\kappa$ of the 
proportional transaction costs tends to $0$.
The asymptotic conditional variance is explicitly connected to 
the initial value of the replicating portfolio, which helps hedgers to
solve the trade-off problem between the hedging error and
the initial cost.
The proposed 
strategy specifies the rebalancing dates that are defined as hitting
times based only on the Black-Scholes Delta and Gamma values.
We have studied the asymptotic mean-squared error as well.
Further, we observed that even under the Black-Scholes model, 
the proposed method improves the well-known strategy of Leland
in a realistic region of parameters.

\end{document}